\documentclass[twocolumn,aps,floatfix,showpacs,prd,longbibliography]{revtex4-2}
\usepackage{graphics, epsfig ,color} 
\usepackage{amsmath,amssymb} 
\usepackage{amsfonts,subfigure}
\usepackage{ulem}
\usepackage{enumerate}
\usepackage[dvipsnames]{xcolor}

\usepackage{xcolor, soul}
\sethlcolor{yellow}

\graphicspath{{figures/}}

\begin{document}
\date{\today}
\title{Frequency-synchronized networks of non-identical pulse-coupled excitable oscillators in the presence of delays proportional to their euclidean distance}
\title{Polychronization as a result of Frequency-synchronized networks}
\title{Self adaptation of networks of non-identical pulse-coupled excitatory and inhibitory oscillators in the presence of distance-related delays to achieve frequency synchronisation}
\author{L. Gil }
\affiliation{Universit\'e C\^ote d'Azur, Institut de Physique de Nice (INPHYNI), France}

\begin{abstract}
 We show that a network of non-identical nodes, with excitable dynamics, pulse-coupled, with coupling delays depending on the Euclidean distance between nodes, is able to adapt the topology of its connections to obtain spike frequency synchronization. The adapted network exhibits remarkable properties: sparse, anti-cluster, necessary presence of a minimum of inhibitory nodes, predominance of connections from inhibitory nodes over those from excitatory nodes and finally spontaneous spatial structuring of the inhibitory projections: the furthest the most intense.
 \end{abstract}

 \pacs{89.75.-k Complex systems - 89.75.Fb Structures and organization of complex systems - 05.45.Xt Synchronization; coupled oscillators -  05.65.+b Self-organized systems}
\maketitle

\section{Introduction}
Phase synchronization of excitable pulse-coupled oscillators in the presence of delays proportional to their distance is a  geometric frustration problem not admitting a solution in general. The basic idea is as follows: for nodes $A$ and $B$ to train each other to spike in phase, the delay $\tau_{AB}$ between them must be an exact multiple of their interspike interval ($ISI$). Similarly, for $B$ and $C$ to spike in phase, $\tau_{BC}$ must be a multiple of $ISI$. But, unless you are in a very particular geometry (such as the one used in \cite{Ko2004,Ko2007,Sethia2008,Zhu2016}) $\tau_{AC}$ is in general not proportional to $ISI$ and therefore the spike of $A$ participates in desynchronizing $C$. Along this argument, the pulse aspect of the coupling, i.e. the existence of an interaction only during a very short time interval compared to $ISI$, is fundamental. Indeed, the further the coupling is from a Dirac distribution, the less the proportionality relation between delay and $ISI$ is constraining.

Given the difficulty of the problem, several approaches have been tried.
Pioneering work \cite{Gerstner1996} deals with a network of identical integrate-and-fire pulse-coupled and excitatory units. The delay is not related to the distance between nodes but to a maximum time beyond which the action of node $j$ on node $i$ is forgotten (i.e. reduces to zero). Two topologies are studied: a fully connected network and a two-dimensional regular mesh with local coupling. The dynamics converges to a frequency synchronized solution, where all nodes have the same $ISI$ without spiking in unison.
In \cite{Bressloff1997}, delays are now clearly associated with the time required for the action potential to propagate along the axon of each neuron. Identical excitatory neurons with an exponentially decreasing coupling with distance, give rise to waves (which implies a global synchronization in frequency but not in phase).
\cite{Ko2002}  investigated the effect of time delays on a set of two-dimensional identical excitatory oscillators. The oscillators are regularly distributed on a square grid and the interactions between oscillators $A$ and $B$ are delayed by an amount proportional to the distance $r_{AB}$ between them. The weights of the connections first decrease as $1/r_{AB}$, then vanish for $r_{AB} > r_{0}$. The oscillators are not pulse-coupled. It is found that distance-dependent time delays induce various patterns including traveling rolls, square-like and rhombus-like patterns, spirals, and targets.
\cite{Atay2003} considered the effects of distributed delays on amplitude death. Oscillators, whose amplitude must be described in order to eventually cancel it, are of Ginzburg-Landau type. They are identical and their dynamics are not excitable. Here the delays are not distance-dependant but are chosen randomly accordingly to a given probability distribution. It is showed that even a small spread in the delay distribution can greatly enlarge the set of parameters for which amplitude death occurs.
The idea of the statistical distribution of delay was then taken up:  to study a standard field model of neural excitatory and inhibitory populations \cite{Hutt2005,Hutt2006}, to investigate the coherent activity patterns in inhibitory, synaptically coupled, bursting Hindmarsh-Rose neurons \cite{Liang2009}, to demonstrate the widespread occurrence of dynamically maintained spike timing sequences in recurrent networks of pulse-coupled spiking neurons with large time delays \cite{Gong2007}.

\cite{Gosak2012} studies the Rulkov mapping in the presence of a delay proportional to the interneuron distance and of a coupling strength proportional to the difference of the fast variables (coupling known as electrical as opposed to synaptic coupling known as pulse-coupling). The neurons are not identical, the dynamics of an isolated neuron is chaotic and the network organization allows a continuous modulation between a scale-free network with dominating long-range connections and a homogeneous network with mostly adjacent neurons connected. A time averaged Kuramoto's order parameter ($R$) is measured. It is found that the most phase synchronized response ($R \simeq 0.4$) is obtained for the intermediate regime where long as well as short-range connections constitute the neural architecture.

All the previous studies we have just described share a common approach: that of providing i) first an excitable dynamics for the nodes and ii) an a priori topology for the network connections, with specific properties such as random, small word, scale free, all to all or sparsely connected, dependence of connection weights on distance etc...Although natural and prolific, this approach leaves a serious doubt about the adequacy of the a priori network topology to the intrinsic properties of the dynamics of the individual nodes. For example, consider a small world network of pulse-coupled neurons with excitable dynamics. These neurons are either excitatory or inhibitory. They also differ in the duration of their refractory period.  Finally, with randomly positioned nodes and delays proportional to the distance between the nodes, the combination of delays seen by each node is absolutely unique. A small-world network is characterised by the presence of a few long-distance connections. But which nodes are best suited to establish long-distance connections? The excitatory ones? the inhibitory ones? the largest or smallest refractory periods? Imposing a network topology without fine-tuning it to the detailed characteristics of each node leaves a lot to chance and is not an optimal way to proceed.

Solving this problem requires to leave the network free to self-adapt to the specificities of each neuron. In line with this finding, \cite{Timms2014} explores both analytically and numerically an ensemble of coupled phase oscillators governed by a Kuramoto-type system of differential equations, where effects of time delay (due to finite signal-propagation speeds) and network plasticity (via dynamic coupling constants) inspired by the Hebbian learning rule, are taken into account. The oscillators are not pulse-coupled and the same neuron can simultaneously project excitatory and inhibitory synapses. In two dimensions, various type of spatiotemporal patterns displaying frequency but not phase synchronisation are then reported. Another approach is to get as close as possible to biological reality. The numerical experiment in \cite{Izhikevich2004} simulates the activity of $10^5$ neurons and $8.5$ $10^6$ synaptic contacts randomly distributed on the surface of a sphere of radius $8$ mm with sub-millisecond time resolution. The neurons interact via both local and long-distance connections. The ratio of excitatory to inhibitory neurons is $4/1$.  Neurons, both excitatory and inhibitory, are not identical and the parameters that describe their dynamics in the absence of coupling are randomly distributed around a mean value. Short-term depression and facilitation are taken into account through the Markram's \cite{Markram1998} phenomenological description of short-term synaptic plasticity. Long-term synaptic plasticity is taken into account through spike-timing dependent plasticity  \cite{Poo1998}. The main result of this numerical experiment is the observation of spontaneous self-organization of neurons into groups and repeatedly generated patterns of activity with millisecond precision of spike timing ( in agreement with experimental observations \cite{Ikegaya2004}). Later noting that the propagation delay between any individual pair of neurons is precise and reproducible
with a sub-millisecond precision \cite{Swadlow1985,Swadlow1994}  and arguing that obtaining and
maintaining such precision can only be understood if the spike-timing is of the highest importance for the brain, Izhikevich introduces the term Polychronization \cite{Izhikevich2006}
to qualify such spiking activity and suggests that they could play a crucial role in the information storage process.

Here we are interested in a network of non-identical excitable oscillators, pulse-coupled, with remote actions, either excitatory or inhibitory, retarded by propagation delays. Our aim is to understand if and how such a network can self-organize to reach a regime of frequency (but not necessarily phase) synchronization? We do not seek to obtain this frequency synchronization by bringing into play biologically realistic mechanisms, but rather approach it as an optimization problem where each node modifies the weight of its incoming connections to best adjust its $ISI$ to an external and common setpoint $ISI_{sp}$. In some ways, we are more interested in the pursued finality and its consequences than in the means to reach it. Our mains results are:
\begin{enumerate}
\item the frequency synchronization requires the mandatory presence of a minimum percentage of inhibitory nodes among excitatory ones.
\item the nodes that spike at the same time and constitute the repeatedly generated patterns of activity with millisecond precision of spike timing reported in \cite{Izhikevich2004,Izhikevich2006,Ikegaya2004} actually form anti-clusters. This means that almost all of the connection weights are associated with inter-pattern links, while the mass of intra-pattern connections is almost vanishing.
\item During the adaptation process, the statistics of the connection weights converge to a lognormal distribution. The weight of outgoing connections from inhibitory nodes  is significantly larger than would be expected if the weights were randomly distributed among the nodes. Those from the inhibitory nodes are on the contrary significantly less numerous. Moreover, we observe the spontaneous occurrence of a spatial structuring where the weight of the outgoing connections is greater and deviates all the more from the random distribution as the distance between the nodes is greater.
\end{enumerate}

The study plan is as follows: First, the excitable dynamics model used will be described and the synchronisation algorithm and its consequences on the network dynamics will be presented. The convergence of the algorithm will then be checked numerically. In a second step we present our results: i) necessity of a minimum percentage of inhibitors, ii) occurrence of death amplitude in the presence of a high percentage of inhibitors, iii)  the formation of anticlusters and iv) spatial distribution of the weights of the connections as a function of the distances and the excitatory-inhibitory nature of the connections. Finally the possible implications of our results to genuine neural networks are discussed.

  \section{The model}
 \subsection{neuronal dynamics}
To model a network of $N$ pulse-coupled excitable oscillators, we use a point process framework \cite{Truccolo2005}. The benefits of such a choice are multiple:
\begin{enumerate}
\item the intrinsically probabilistic nature of the dynamics. We obtain a Poisson's distribution of inter-spikes interval for an isolated neuron without any effort.  
\item the perfect control of the dynamics of a neuron. The temporal evolution of an isolated neuron requires the integration of neither a dynamic system nor the computation of a nonlinear mapping but just corresponds to a shift in the state space.
\item and above all a remarkable efficiency and speed of execution.  The algorithm does not converge all the time, and even when it does, it can take several tens of millions of integration steps, hence the need to go fast.
\end{enumerate}
The drawbacks are the consequence of the advantages: the dynamics of an isolated neuron is highly schematized, especially compared to the diversity of possible behaviors and to the precise modeling that could be done \cite{Izhikevich2003}.

The state of neuron $i$ at time t ($t \in \mathbb{N}$) is described by the variable $S_{i}(t)$ which takes discrete values in $[-T^{r}_{i},T^{s}]$. $T^{s}$ and $T^{r}_{i}$ are integer values representing respectively the spike and the refractory durations. The neurons are not identical because they can differ by the duration of their refractory period $T^{r}_{i}$. The dynamics of $S_{i}$ is composed by an alternation of a deterministic and a stochastic part. The deterministic part starts at time $t^*$ whenever $S_{i}(t^*)=T^{s}$ and continues with

\begin{widetext}
\begin{tabular}{|c|c|c|c|c|c|c|c|c|c|}
  \hline
  $t^*+...$ &$0$ &   $1$ & ... &  $T^{s}-1$ &   $T^{s}$ &   $T^{s}+1$ & ... &   $T^{s}+T^{r}_{i}-1$ &   $T^{s}+T^{r}_{i}$ \\
  \hline
  $S$ & $T^{s}$ & $T^{s}-1$ & ... & $1$ &$ -1$ & $-2$ & ... & $-T^{r}_{i}$ & $0$ \\
  \hline
\end{tabular}
\end{widetext}

Note that during this deterministic sequence, $S_{i}$ jumps from $+1$ to $-1$ without passing through $0$. This is because we reserve $S_{i}= 0$  to describe the rest state,  the one reached after the refractory period. The stochastic part starts at time $t^{rest}$ whenever $S_{i}(t^{rest})=0$ and is involved in the determination of the next state $S_{i}(t^{rest}+1)$
\begin{equation}
S_{i}(t^{rest})=0 \Longrightarrow S_{i}(t^{rest}+1)=
\left\{
\begin{array}{lcl}
T^{s} &{\rm with \, prob} & p_{i}(t^{rest})
\cr
0 & {\rm "}  & 1-p_{i}(t^{rest})
\end{array}
\right.
\label{PointProcess}
\end{equation}
with
\begin{equation}
p_{i}(t)={\cal R} {\Big [} p_{0}+ a \displaystyle{\sum_{j=1}^{N}} D_{j} W_{ij} H(S_{j}(t-\tau_{ij})) {\Big ]}
\end{equation}
where 
\begin{equation}
R(x)=\left\{
\begin{array}{ll}
0 & \rm{if} \,\,\, x \le 0 
\cr
x & \rm{if} \,\,\, 0 \le x \le 1
\cr
1 & \rm{if} \,\,\, x \ge 1
\end{array}
\right.
\qquad
H(n)=\left\{
\begin{array}{ll}
 1 & \rm{if} \,\,\, n>0
 \cr
 0 & \rm{otherwise} 
 \end{array}
\right.
\end{equation}
$p_{0} \in [0,1]$ and $a \ge 0$ are constant parameters, $D_{j}=\pm 1$ depending on whether $j$ is excitatory or inhibitory, $W_{ij} \ge 0$ is the strength of the connection from $j$ to $i$ and $\tau_{ij}$ is their propagation delay proportional to their Euclidean distance. The role of the function $R$ is to guarantee that $p_{i}$ is a probability, that is a positive number in $[0,1]$. The pulse-coupled character of the dynamics is modeled by the function $H$ which takes non-zero values only when the neighbors spike at the right time.

When the neuron chains spikes without discontinuity, its dynamics is periodic and the inter-spike interval ($ISI$) reaches its minimum value $\Delta=T^{s}+T^{r}_{i}+1$. In our simulations, we use $p_{0}=0.001$ such that the average $ISI$ in absence of coupling ($a=0$) is about $10^3$ time steps.

 \subsection{network geometry}
In line with our objectives, the network is free to adapt as it wishes since it is all to all connected and that the weights of the connections $W_{ij}$ can evolve without constraints between $[0,+\infty]$. On the other hand, the spatial positions of the nodes and consequently the propagation delays are determined once and for all at the beginning of the optimization process. In what follows, we discuss this initial distribution of positions that we want to be both random but with a well-defined smaller distance between neighbors \cite{Wassle1978}.

\begin{figure}
\resizebox{0.40\textwidth}{!}{
\includegraphics[]{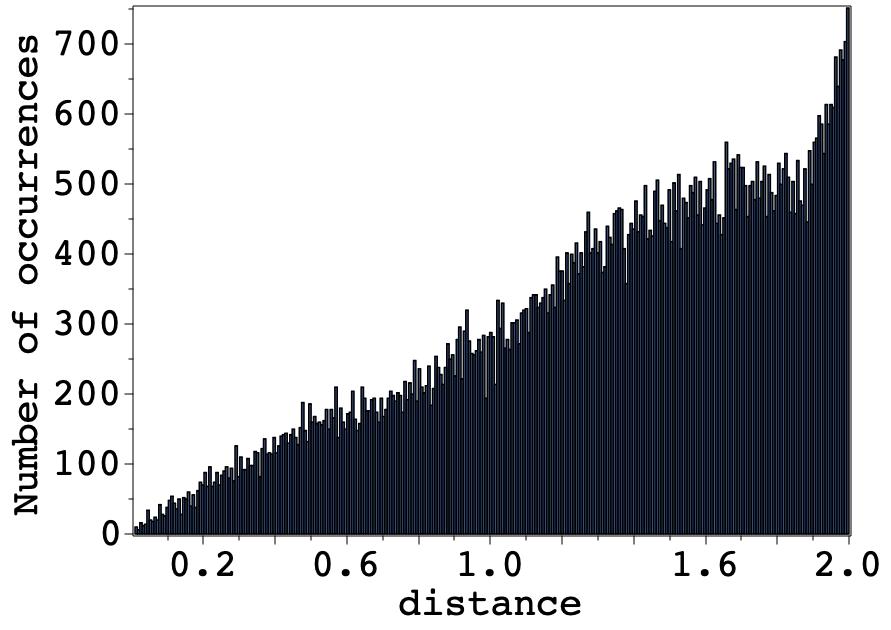}
}
\caption{Histogram of the interneurons distances before ajustement with $300$ bins (vanishing distances are not taken into account). The network has 300 nodes randomly distributed on a sphere of unit radius .}
\label{fig01}
\end{figure}

In a first step, $N$ neurons are randomly distributed on the surface of a sphere of radius $R=1$. The interneuron distances vary between $0$ and $2R=2.0$ and their initial distribution is shown in fig.\ref{fig01}.

\begin{figure}
\resizebox{0.40\textwidth}{!}{
\includegraphics[]{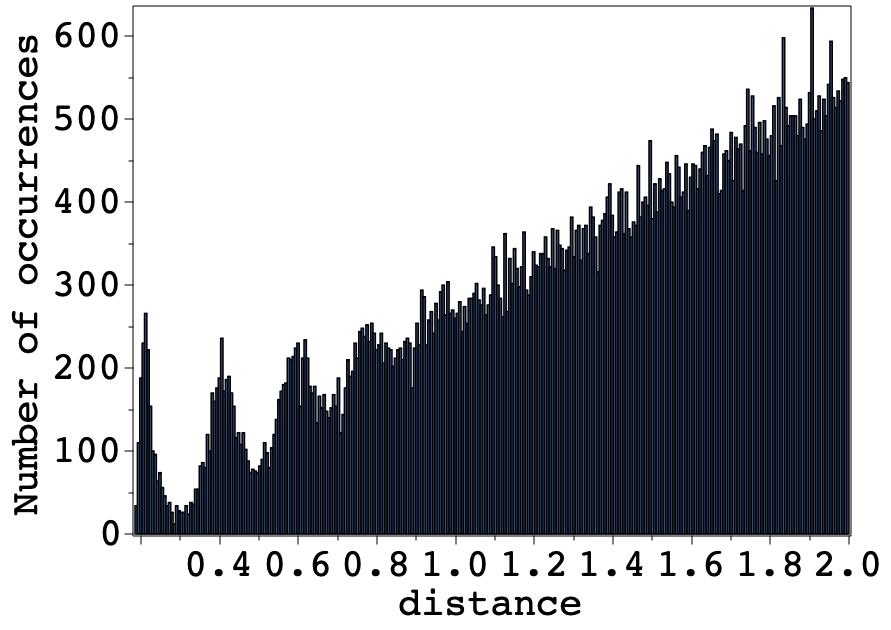}
}
\caption{Histogram of the interneurons distances after ajustement. The first peak (the most left-handed) in the distribution is associated with  $\displaystyle{\min_{ j}} (r_{ij}) $, i.e. the mesh of the hexagonal network. The ratio between the height of this peak and its width at half height defines the quality factor.}
\label{fig02}
\end{figure}

In a second step the spatial distribution of the nodes onto the surface is regularized in order to homogenize their surface density. This adjustment is achieved by subjecting the node $i$ to repulsive $\sum_{j} 1/r_{ij}$ interactions. The repulsive forces are applied until the quality factor of the $\displaystyle{\min_{ j}} (r_{ij})$ distribution is equal to $30$ \cite{Wassle1978}. In the end, the nodes form an almost hexagonal network (with mesh $d_{hex} \simeq < \displaystyle{\min_{j}} (r_{ij}) >$), with many penta-hepta topological defects (fig.\ref{fig02}). 

As the simulation is time discretized, all the delays $\tau_{ij}$ are expressed as integer unit of $cdt$ the distance traveled by the information during a unit of time:
 \begin{equation}
 \tau_{ij}=\left[ {{r_{ij}}\over{cdt}}\right]
 \end{equation}
where $[\,]$ stands for the integer value. An important parameter is then the number of time steps necessary to transmit the information from one node to its nearest neighbor. 
This number is equal to $\tau_{min}=d_{hex} / cdt$. The maximum distance being $2R=2$, the state of all the neurons must be stored in memory over a duration of $2 \tau_{min} / d_{hex}$ time steps. Therefore, for an economical management of the memory it is better to use a small value of $\tau_{min}$ (in most of our simulations we used $\tau_{min}=3$).


\section{Algorithm}
There are no strict and rigorous rules leading to the choice of the algorithm used. Rather, it is the result of a set of general considerations, analogies and heuristic arguments that we present below. Ultimately, the main rationale is that it effectively leads to synchronized solutions.

\begin{enumerate}
\item We have deliberately chosen not to use a central control capable of accepting or rejecting a solution based on a global computation. The reason is that this kind of approach quickly becomes impractical with increasing $N$. On the contrary, we opted for a local, scalable and parallelizable approach. 

\item Following H.A. Simons' ideas in his famous paper "Architecture of the complexity" \cite{Simon1962}, the nodes of the network were imposed to be unable to perform complicated mathematical computations (such as, for example, gradient computations or predictions). We just expect each oscillator to be able to estimate its $ISI$ and to compare it with the setpoint $ISI_{sp}$.

\item We assume that the incoming weight adjustment is not done systematically at each time step but only when the node has just spiked.

\item A node that has just tested a new local weight configuration but which ultimately does not adopt it, cannot force the rest of the network to return to its initial state configuration. This would require too much effort in terms of storage and transport of information. The node that did the test must continue on its way. Optimization must be done on the fly.

\item The modification of the incoming connections of node $A$ has a direct effect on its spike frequency. On the contrary, the effect of modifying its outgoing connections is obviously more indirect: when node $A$ acts on the incoming connections of its neighbours, then their spike frequencies are modified and may act in return on the spike frequency of $A$. Both approaches are possible but we will limit ourselves in the algorithm to the most efficient one, i.e. the modification of incoming connections only 

\item We have chosen not to impose any a priori structure on the connection network. Each node is connected with all the others but the weights of the connections evolves without constraint, can vanish or, on the contrary, grow indefinitely. This is a very expensive choice in terms of computating time but which is absolutely necessary to let the network freely choose its own topology. 

\end{enumerate}
If, at time $t$, node $i$ does not spike, then its incoming connections do not change.  Now if it spikes at time $t$, then this node starts its remodelling activity by estimating the elapsed time interval $ISI$ between its last two spikes.  Let $j$ be another node of the network ($j \ne i$) connected to $i$ through $W_{ij}$. If the last spike of $j$ took place at a time different from $t-\tau_{ij}$, then $i$ does not perceive any synaptic potential from $j$. It is then useless for itself to maintain the incoming connection $W_{ij}$ and
\begin{equation}
W_{ij}(t+1)=W_{ij}(t)\left(1-b\right)
\end{equation}
where $b$ is a small positif real. On the contrary, if $j$ spiked at $t-\tau_{ij}$ then its weight contribution is changed accordingly to
\begin{equation}
W_{ij}(t+1)=max \Big{(} 0,W_{ij}(t)+\alpha \xi D_{j} \left(ISI-ISI_{sp}\right) \Big{)}
\label{Wdynamique}
\end{equation}
where $\xi \in [0,1]$ is a random uniformly distributed variable and $ \alpha \ge 0 $ stands for the modification amplitude. Interpretation of eq.(\ref{Wdynamique}) is straightforward: if $i$ detects that its $ISI$ is higher than the setpoint (i.e. $\left(ISI-ISI_{sp}\right) > 0$), the incoming connections associated with inhibitory nodes ($D_{j}=-1$) will be decreased while those associated with excitatory nodes ($D_{j}=+1$) will be increased. As a result, $W_{ij}$ and the probability for $i$ to spike are increased. 
Conversely, when $ISI < ISI_{sp}$, the same dynamics eq.(\ref{Wdynamique}) leads to a decrease in the spike probability. Note that more sophisticated expressions can be considered for the weight change, but Eq.\ref{Wdynamique} can be understood as the unique linearization in the neighborhood of $ISI \simeq ISI_{sp}$ of any mechanim imposing frequency synchronisation.

It is important to realize that the algorithm is of the greedy type.  Although the evolution of $W_{ij}$ (eq.\ref{Wdynamique}) imposes without any doubt that the ISI of node $i$ will get closer to the setpoint, the simultaneous global convergence of all the nodes is absolutely not guaranteed: the convergence of a node can be done at the expense of another one.

\begin{figure}
\resizebox{0.40\textwidth}{!}{
\includegraphics[]{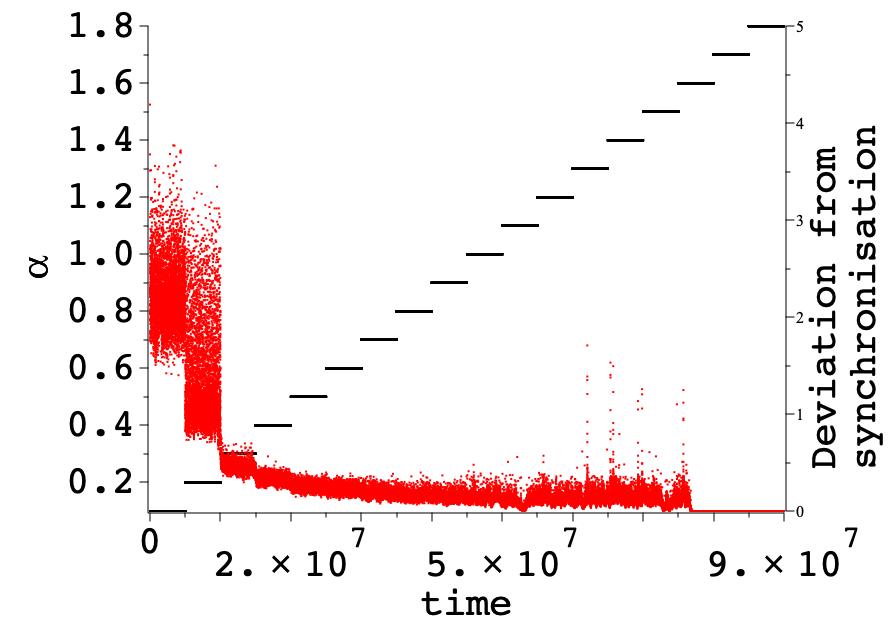}
}
\caption{Imposed time evolution of $\alpha$ (left axis, in black) and $G_{s}=\sum_{i=1}^{N} (ISI_{i}-ISI_{sp})^2$ (right axis, in red) along the optimization process. The network consists in $300$ nodes: for $T_{r}=38$  there are  $88$ excitatory and $16$ inhibitory nodes, for $T_{r}=39$, $96$ and $10$ and for $T_{r}=40$, $71$ and $19$. $T_{s}=3$ and $a=4$. The $ISI$ setpoint is set at $45$.}
\label{fig04}
\end{figure}

Fig.\ref{fig04} represents a typical time evolution of the global deviation $G_{s}=\sum_{i=1}^{N} (ISI_{i}-ISI_{sp})^2$ along the optimization process. While $\alpha$ is gradually increased by steps of $0.1$, we observe a decrease of $G_{s}$ to zero indicating that the system does evolve globally towards a frequency synchronization. However, this convergence is far from being uniform and takes rather the aspect of an avalanche dynamic where the local optimization of a node can provoke a cascade of events at the network level. Finaly, when the global synchronization is reached, the network dynamics stops and the network does not evolve anymore.

Randomness is present in the dynamics through $p_{i}(t)$ (eq.\ref{PointProcess}) and $\xi$ (eq.\ref{Wdynamique}) and the initial geometrical distribution of the nodes. To investigate these effects, we perform two types of numerical experiments. All the simulations have in common the same parameter values ($a$, $cdt$, $T_{s}$ and $ISI$ setpoint), the same initial $W_{ij}$ values and  they share the same distribution of refractory periods and excitatory/inhibitory ratios: for $T_{r}=38$, $80/20$, for $T_{r}=39$, $79/21$ and for $T_{r}=40$ $81/19$ (we introduce the notation $[[38,80/20],[39,79/21],[40,81/19]]$ to designate such a configuration).  On the other hand, the two groups differ by their initial distribution of the position of the nodes. The simulations of the first group ($20$ simulations) use a strictly identical geometrical distribution such that the origin of randomness is limited to $p_{i}(t)$ and $\xi$. We observe that the convergence toward a frequency synchronisation regime is achieved for $\alpha > \alpha_{c}$ where $\alpha_{c}$ varies from one experiment to another with $\alpha_{c} \in [0.8,1.2]$. Averaging over the $20$ experiments, we found $ <\alpha_{c}>= 1.0 \pm 0.1$. Each of the simulations of the second group ($10$ simulations) uses its own, randomly generated,  geometrical configuration. We found $\alpha_{c} \in [0.90,1.60]$ with $<\alpha_{c}>=1.1\pm 0.2$. Thus, we can see that i) the two types of measures are consistent with each other, ii) and that the random distribution of node positions is an important source of fluctuations. Therefore, in what follows, each optimization process will be associated with a random draw of the position of the nodes.

\section{Results}
\subsection{Spatio-temporal dynamics at convergence}

\begin{figure}
\resizebox{0.40\textwidth}{!}{
\includegraphics[]{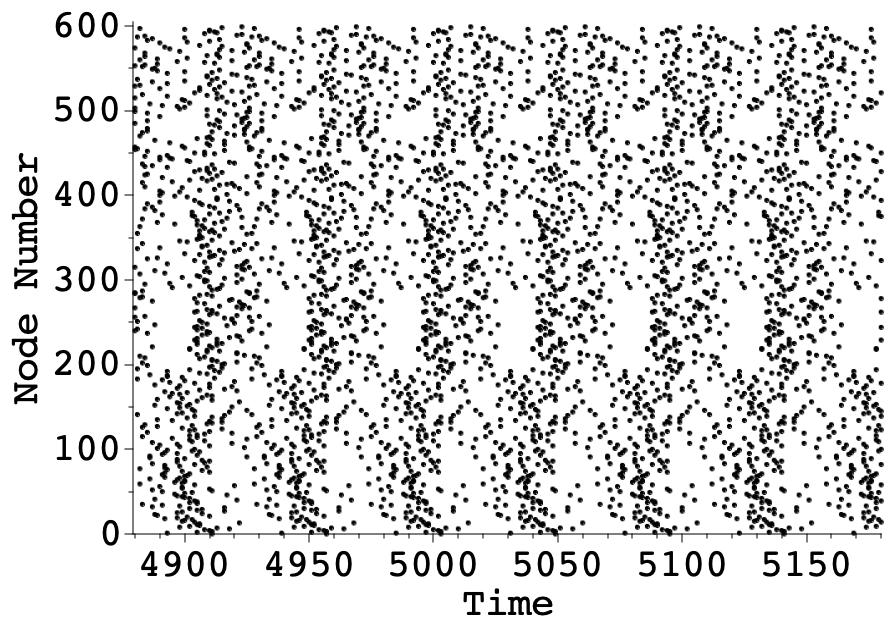}
}
\caption{Raster plot of the network activity. The configuration is $[[38,136/53],[39,157/53],[40,170/31]]$ and involved $600$ nodes. $T_{s}=3$, $p_{0}=0.001$, $b=0.01$, $a=4$ and $ISI_{sp}=46$. The figure corresponds to the spatiotemporal dynamics after convergence of the optimization process.}
\label{fig05}
\end{figure}

\begin{figure}
\resizebox{0.40\textwidth}{!}{
\includegraphics[]{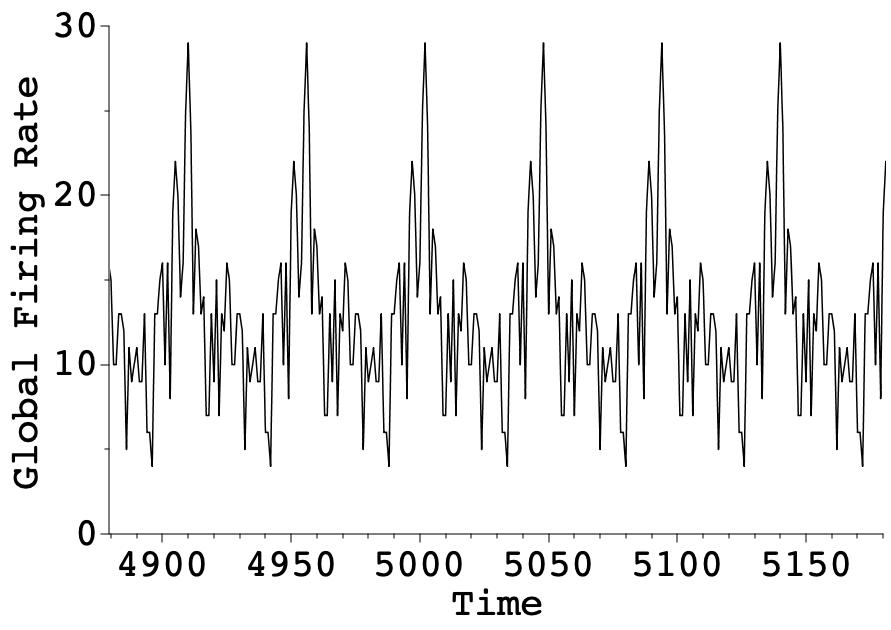}
}
\caption{Time evolution of the global firing rate associated with the raster plot in fig.\ref{fig05}.}
\label{fig06}
\end{figure}

\begin{figure}
\resizebox{0.40\textwidth}{!}{
\includegraphics[]{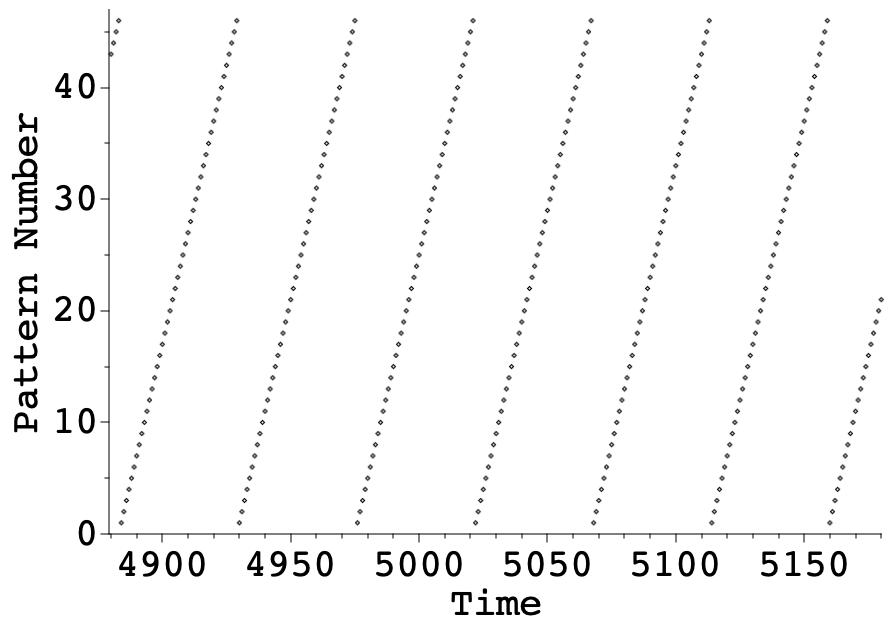}
}
\caption{Time evolution of the spatio-temporal dynamics in the space of the patterns. On the vertical axis, the numbers $1$ to $46$ stand for the patterns $P_{1}$, $P_{2}$...$P_{ISI_{sp}}$ observed at the convergence of the optimization process in  fig.\ref{fig05} and fig.\ref{fig06}.}
\label{fig07}
\end{figure}

\begin{figure}
\resizebox{0.40\textwidth}{!}{
\includegraphics[]{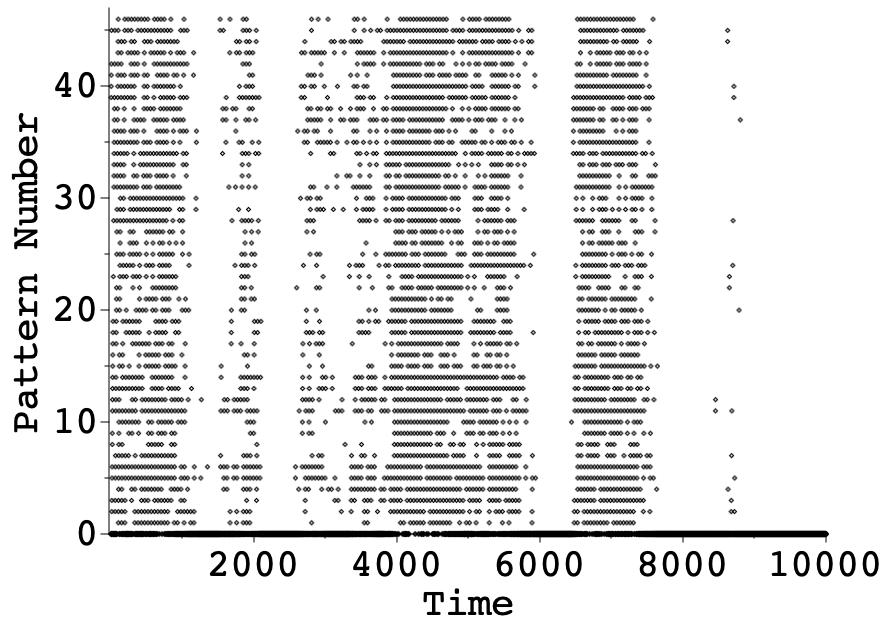}
}
\caption{Same as fig.\ref{fig07} but now $p_{0}=0.04$ such that the dynamics is strongly disrupted. As before, the numbers $1$ to $46$ on the vertical axis stand for the patterns ${\cal D}=P_{1}$, $P_{2}$...$P_{ISI_{sp}}$ but now $0$ is associated with any patterns that is not in the list ${\cal D}$. Pay attention to the difference in the horizontal scales: the one in fig.\ref{fig07} spans only a few $ISI_{sp}$ while here it corresponds to more than $200$.}
\label{fig08}
\end{figure}

At convergence, the spatio-temporal dynamics is characterized by the periodic succession of node patterns ${\cal D}=$ $P_{1}$, $P_{2}$...$P_{ISI_{sp}}$ where $ISI_{sp}$ is the imposed inter-spike interval setpoint (fig.\ref{fig05}). A pattern is constituted by the set of all nodes that spike at the same time. As the number of nodes varies from one pattern to another, the global firing rate oscillates periodically in time with the period $ISI_{sp}$ (fig.\ref{fig06}). The patterns in the sequence ${\cal D}$ are 2 by 2 disjoint and their gathering constitutes the total set of nodes of the network. Therefore they form a partition of the set of nodes.
Fig.\ref{fig07} and fig.\ref{fig08} show typical temporal evolutions of the dynamics in the space of the patterns. On the vertical axis, the zero corresponds to any pattern that is not in the list ${\cal D}=$$P_{1}$, $P_{2}$...$P_{ISI_{sp}}$. Fig.\ref{fig07} is the regular and periodic dynamics obtained after convergence of the optimization process. Fig.\ref{fig08} is obtained by freezing the dynamics of the network corresponding to fig.\ref{fig07} (i.e. $W_{ij}$ are constant) and by increasing the background noise ($p_{o}=0.04$). The global dynamics is found to be intermittent with phases of locking on the periodic solution at convergence, interspersed by episodes of more or less long stall with a complex dynamics.

\subsection{Mandatory presence of inhibitory nodes and amplitude death}
The importance of inhibitory mechanisms for generation of cortical rhythms is now well established \cite{Bibbig2002,Mann2007}: Synaptic inhibition is known to balance excitation and control the precise timing of spike generation. Synaptic inhibition itself can be synchronized by way of interactions within networks of inhibitory and excitatory neurons. 
It is therefore expected that our model also proves that frequency synchronization is only possible in the presence of a minimum number of inhibitory nodes.

Each node being associated with a specific refractory period $T_{r}$, we should normally characterize a given network by its statistical distribution of $T_{r}$.  Nevertheless, for the sake of simplicity, we have concretely limited ourselves to $3$ distinct values (typically $T_{r} \in \left[38,40\right]$). Tests with up to $5$ values have been performed to check that this limitation was not relevant. 
The spike duration $T_{s}$ being the same for all nodes, the setpoint for the interval between 2 spikes $ISI_{sp}$ cannot be less than $\Delta_{min}=T_{{r}_{min}}+T_{s}+1$ because our model (eq.\ref{PointProcess}) does not contain any mechanism capable of reducing the refractory period. On the other hand, it seems possible to impose an $ISI_{sp}$  greater than $\Delta_{max}=T_{{r}_{max}}+T_{s}+1$ because one expects the inhibitory neurons to cooperate to prohibit the spike over a duration longer than $T_{{r}_{max}}$.  Typically we impose 
either $ISI_{sp}=\Delta_{max}+1$ or $ISI_{sp}=\Delta_{max}+2$. Control simulations with $ISI_{sp}=\Delta_{max}+5$ have been successfully performed. However, for even larger values, numerical convergence problems have been encountered.

\begin{figure}
\resizebox{0.40\textwidth}{!}{
\includegraphics[]{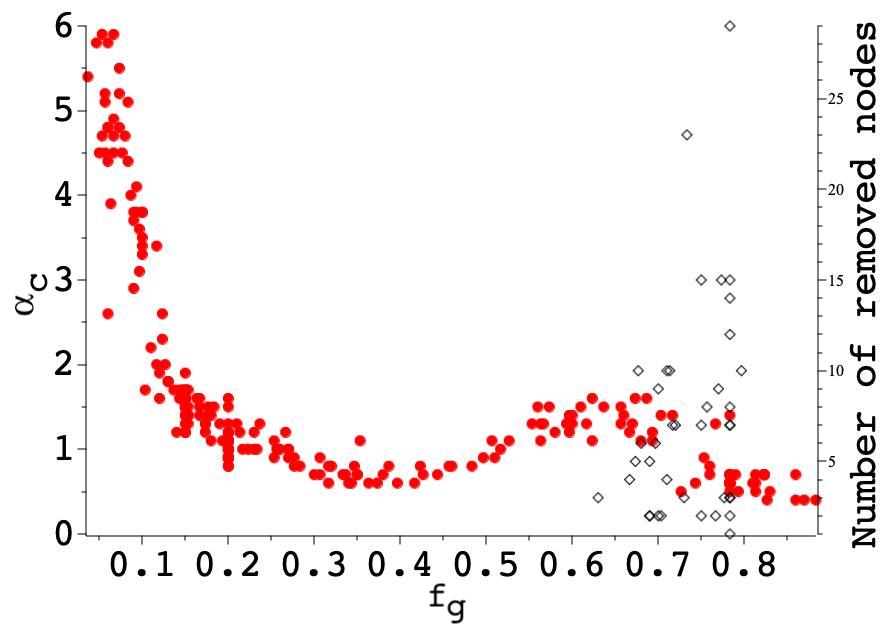}
}
\caption{The network consists of 300 nodes whose initial positions are randomly chosen on a sphere. The refractory period of each node is randomly chosen among the 3 values $38$, $39$ and $40$ and its inhibitory/excitatory character is determined by drawing with a probability $f_{g}$ (fraction of inhibitors). The $ISI$ setpoint is $45$. The red points (left axis, solid discs) stand for $(f_{g},\alpha_{c})$. The black ones (hollow diamonds) correspond to $(f_{g},n_{death})$, where $n_{death}$ is the number of nodes that have ceased to spike under the pressure of the inhibitory nodes along the optimisation process}. 
\label{fig09}
\end{figure}

We have conducted no less than $300$ numerical experiments (fig.\ref{fig09}). For each simulation, the initial position of the nodes is randomly generated. Then for each node, its value of $T_{r}$ is chosen randomly and uniformly between the $3$ values $38$, $39$, and $40$. Finally the excitatory or inhibitory action of the node is randomly drawn: with a probability $f_{g}$ the node is inhibitory, with a proba $1-f_{g}$ it is excitatory. $f_{g}$ changes with the experiments inside $\left[0.05,0.95\right]$. For each simulation, $\alpha$ is increased in steps of $0.1$ until a critical value $\alpha_{c}$ is reached for which a frequency synchronization regime is established. Value of $\alpha$ higher than $6.0$ have not been investigated. Red points in fig.\ref{fig09} represents the set of $(f_{g},\alpha_{c})$ points. When several $\alpha_{c}$ are associated to the same value of $f_{g}$, it is the highest value of $\alpha_c$ that counts, the one that ensures the convergence towards the frequency synchronization whatever the initial geometry of the nodes and the optimization path taken.
For $f_{g} \simeq 0$, the plot suggests a divergence of $\alpha_{c}$ associated with the impossibility of a global frequency synchronization in the absence of inhibitory nodes. For $f_{g} \simeq 1$, we observe the spontaneous death of a certain number $n_{death}$ of nodes during the optimization process. At a given moment, under the action of their inhibitory connections, these nodes were unable to spike.  And since a node that does not spike cannot change its incoming connections, the situation persists as long as the neighborhood action goes on.

\subsection{Anti-clusters structuring}
\label{Anti}
For two nodes $A$ and $B$ to train each other to spike in phase, the delay $\tau_{AB}$ between them must be an exact multiple of the setpoint $ISI_{sp}$. 
Consequently, we expect and observe two very distinct operating regimes depending on whether the maximum delay between $2$ nodes of the network ($ 2 / cdt$) is less or greater than $ISI_{sp}$.

We introduce
\begin{equation}
R(\left\{ W \right\})=
{{\displaystyle{\sum_{P_{\mu} \in {\cal D}}} \,\,\, \displaystyle{ \sum_{i \in P_{\mu},j \in P_{\mu}}} W_{ij}}
\over
{\displaystyle{\sum_{P_{\mu} \in {\cal D}}} \,\,\, \displaystyle{\sum_{P_{\mu' \ne \mu} \in {\cal D}}} \,\,\, \displaystyle{ \sum_{i \in P_{\mu},j \in P_{\mu'}}} W_{ij}}}
\label{eqR}
\end{equation}
which, for a given configuration $\left\{ W \right\}$, stands for the ratio between the total weight of the internal connections to each pattern $P_{\mu}$ and the total weight of the connections between two distinct patterns $P_{\mu}$ and $P_{\mu' \ne \mu}$. We compute $R(\left\{ W_{cvg} \right\})$ where $\left\{ W_{cvg} \right\}$ is the configuration network at the convergence of the optimization process. We compare the previous result with the distribution of $R(\left\{ W_{rand} \right\})$ where $\left\{ W_{rand} \right\}$ are derived from $\left\{ W_{cvg} \right\}$ by randomly redistributing its weights  among the nodes of the network.

\begin{figure}
\resizebox{0.40\textwidth}{!}{
\includegraphics[]{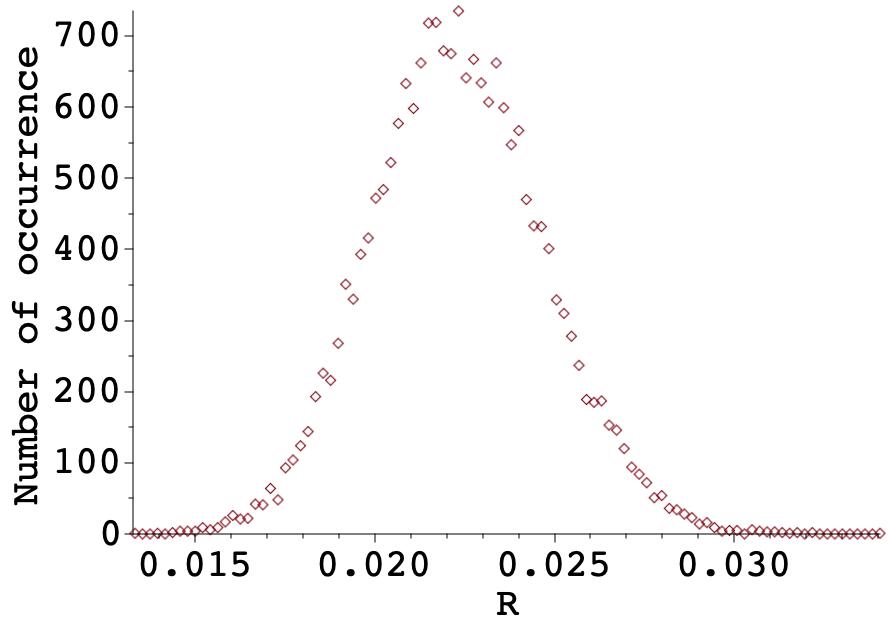}
}
\caption{Histogram of $R(\left\{ W_{rand} \right\})$ defined in eq.\ref{eqR}. We made $10000$ random draws and the histogram has $100$ bins.The network consists of 300 nodes whose initial positions are randomly chosen on a sphere. The refractory period of each node is randomly chosen among the 3 values $38$, $39$ and $40$ and the fraction of inhibitors $f_{g}=0.15$. The $ISI$ setpoint is $45$ while the delay between two diametrically opposed nodes is $2 / cdt = 33 < ISI_{sp}$. We find $<R( \left\{ W_{rand} \right\} ) > = 0.022 \pm 0.0025$ which implies that $R(\left\{ W_{cvg} \right\})$ at convergence deviates from the mean value by more than $9.4$ standard deviation.}. 
\label{fig10}
\end{figure}

We first consider the situation where $ 2 / cdt < ISI_{sp} $ that corresponds to a "small" network where all nodes are within one inter-spike interval of each other. This regime correspond to the vast majority of our investigations because it is the most interesting and the surprising situation.  Fig.\ref{fig10} is a histogram of the values of $R(\left\{ W_{rand} \right\})$ obtained after $10000$ draws of the random configuration $\left\{ W_{rand} \right\}$. While $<R( \left\{ W_{rand} \right\} )>\simeq 0.022$ with a standard deviation of $0.0025$, the measured value at convergence is $R(\left\{ W_{cvg} \right\})=1.9 \, 10^{-7}$, significantly smaller. It thus deviates from the random distribution by more than $9$ standard deviations, which rules out any coincidence: therefore the patterns $P_{\mu} \in {\cal D}$ are characterized by a very strong anti-cluster structuring. 

\begin{figure}
\resizebox{0.40\textwidth}{!}{
\includegraphics[]{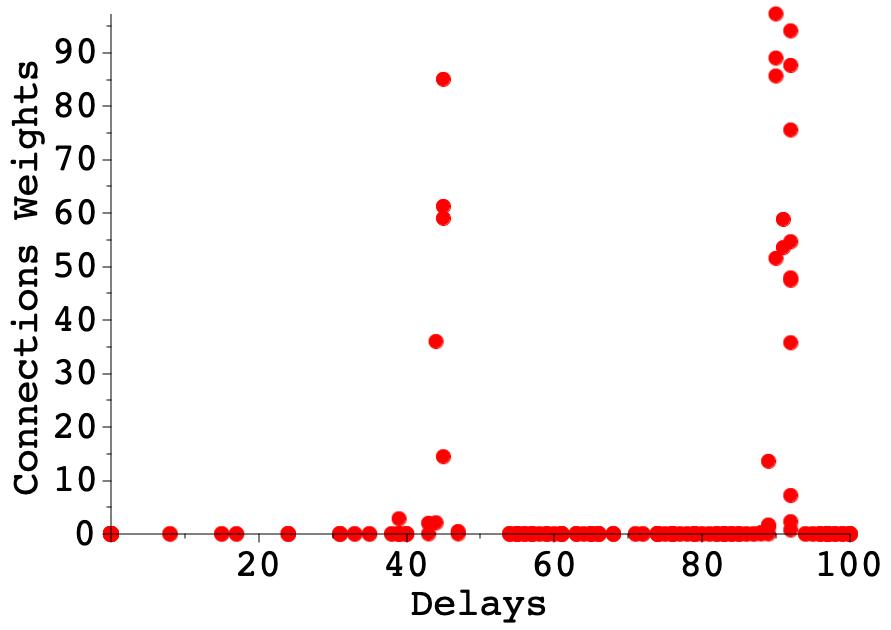}
}
\caption{The plot shows the set of points $\left(\tau_{ij},W_{ij}\right)$ where $i$ and $j$ belong to the same pattern $P_{\mu} \in {\cal D}$. The networks has $600$ nodes, $T_{s}=3$, $a=4$ and $p_{0}=0.001$. Their refractory periods are not identical and vary between $38$ and $40$. The ISI setpoint is fixed at $46$ and $2 /cdt=100$. The fraction of inhibitory nodes is $20 \%$. The first maximum is located at $46$ ($=ISI_{sp}$) and the second at $92$.} 
\label{fig15}
\end{figure}

In the case of a network with $2 /cdt = 100 > ISI_{sp}$, the situation is completely changed. In such a "large" network, each node can be linked to several distinct nodes shifted by exactly one $ISI_{sp}$. Then $R(\left\{ W_{cvg} \right\})$ is no longer almost zero, but on the contrary is measured to be almost one standard deviation higher than $<R(\left\{ W_{rand} \right\})> $ (not shown).  The anti-cluster structure is in competition with the connections between nodes belonging to the same pattern and is clearly less predominant. Fig.\ref{fig15} shows the connection weights repartition $W_{ij}$ versus the delay $\tau_{ij}$ when both $i$ and $j$ belongs to the same given pattern $P_{\mu}$ (randomly chosen in the ${\cal D}$ sequence). We clearly observe that only internal connections with a delay equal to $ISI_{sp}$ ot $2$ $ISI_{sp}$ are not vanishing. 

\subsection{Network sparseness}
The Gini coefficient is a real number, between $0$ and $1$, that measures the rate of inequality of the distribution of a variable. It was originally developed in economics to measure the income inequality of a country's population. Applied to the case of connection weights, a null value of this coefficient would correspond to the homogeneous distribution of the mass, i.e. to the case where all $W_{ij}$ are equal. On the contrary, a coefficient equal to $1$ would mean that all the weights are zero, except for one and only one. For values of $f_{g} \simeq 0.2$ and the number $N$ between $100$ to $600$ of nodes, we find a staggering value of $0.95$ indicating that the optimized networks are particularly sparse with a very large majority of connections reduced to zero coexisting with a very few number of very massive connections.  
\begin{figure}
\resizebox{0.40\textwidth}{!}{
\includegraphics[]{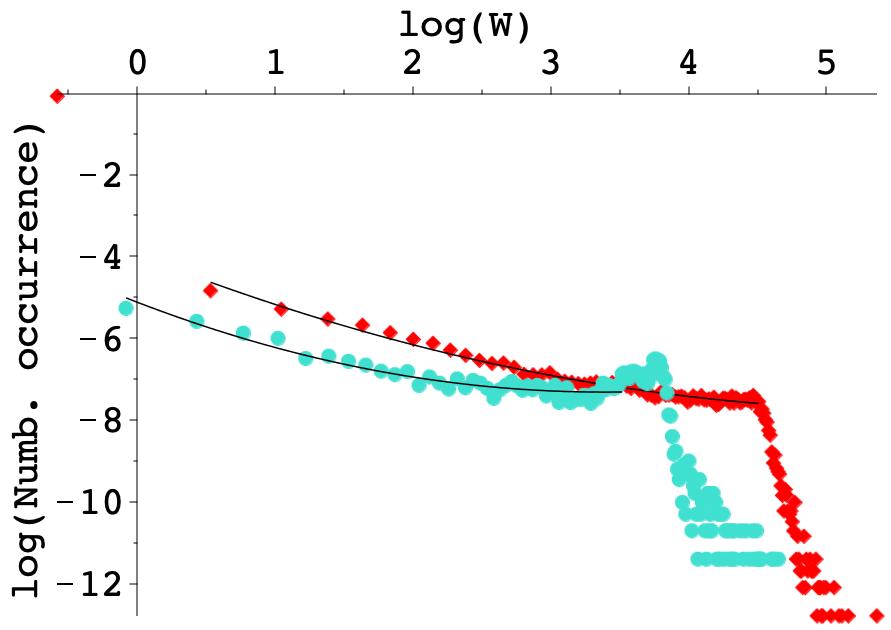}
}
\caption{Histogram with $200$ bins of the weights of the connections $W_{ij}$ in log-log scales. We integrated the results obtained for $5$ distinct networks with $f_{g}=0.2$ at convergence of the optimization process. The turquoise cercles correspond to networks with $300$ nodes while the red diamonds are associated with network with $600$ nodes. Continuous lines are quadratic fits compatible with lognormal distributions.}. 
\label{fig11}
\end{figure}
Fig.\ref{fig11} shows a typical histogram of the connexion weights $W_{ij}$ in log-log scales. 

\subsection{Predominance of projections from inhibitory nodes}
Here we focus on the global masses of the network connections according to the excitatory or inhibitory nature of the nodes of departure and arrival. We introduce
\begin{equation}
P_{++}(\left\{ W \right\})=\displaystyle{\sum_{\scriptsize{\begin{array}{l} \left\{i \in [1,N] \vert D_{i}=+1 \right\} \cr \left\{j \in [1,N] \vert D_{j}=+1 \right\} \end{array}}}} \!\!\!\! \!\!\!\! \!\!\!\!\!\!\!\! W_{ij} 
\label{Pxydefinition}
\end{equation}
where  $P_{++}(\left\{ W \right\})$ stands for the total mass of the excitatory $\leftarrow$ excitatory connections for the network configuration $\left\{W\right\}$. We define in the same way the other masses $P_{+-}$, $P_{-+}$ and $P_{--}$.

The numerical values of the above quantities at the convergence of the optimization process do not have any meaning in themselves. Neither do their ratios since they depend on $f_{g}$. So we will proceed as for the demonstration of the anti-cluster structure in paragraph \ref{Anti}, by comparing $P_{..} \left( \left\{ W_{cvg} \right\} \right)$  with the distribution of $P_{..}(\left\{ W_{rand} \right\})$ where $\left\{ W_{rand} \right\}$ are derived from $\left\{ W_{cvg} \right\}$ by randomly redistributing its weights  among the nodes of the network. The results are displayed in fig.\ref{fig12.5}. Since they differ from the mean values by several standard deviations, they are highly significant from a statistical point of view.  They clearly shows a very net deficit in the mass of the connections from excitatory nodes, to the benefit of a substantial excess in the mass of the connections from inhibitory nodes. 
 \begin{figure}
\resizebox{0.50\textwidth}{!}{
\begin{tabular}{|c|c|c|c|c|}
  \hline
   $ .. $ & $P_{xy}(\left\{ W_{cvg} \right\})$  &   $<P_{xy}(\left\{ W_{rand} \right\})>$ & standard deviation $ \sigma$ & ${{P_{xy}(\left\{ W_{cvg} \right\})-<P_{xy}(\left\{ W_{rand} \right\})>}\over{\sigma}}$  \\
  \hline 
  $++$ & 557485 & 623876 & 3690 & -18.0 \\
 \hline 
  $-+$ & 141147 & 159531 & 2792 & -6.6 \\
  \hline 
  $+-$ & 224183 &159530 & 2782 & 23.2  \\
 \hline 
  $--$ & 60525 & 40403 & 1516 & 13.3 \\
  \hline
\end{tabular}
}
\caption {Same configuration as in fig.\ref{fig05}. $P_{..}$ are defined in eq.\ref{Pxydefinition}. The rightmost column shows the difference between the measurements at convergence and the mean value in units of standard deviation. Undoubtedly, a large part of the mass has been allocated to the outgoing connections from the inhibitory nodes at the expense of the outgoing connections from the excitatory ones. }
\label{fig12.5}
\end{figure}

\subsection{Spatial distribution of the connexion weights}
\begin{figure}
\resizebox{0.40\textwidth}{!}{
\includegraphics[]{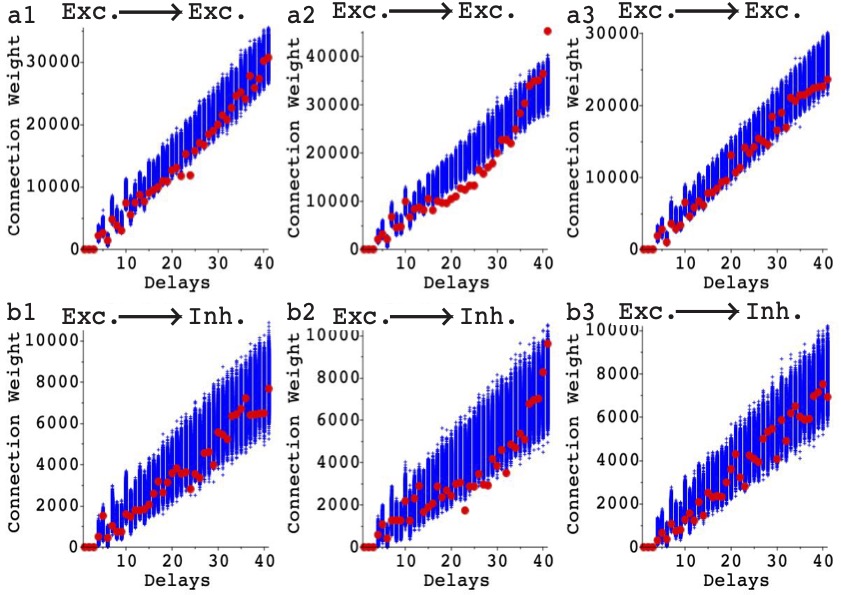}
}
\caption {The first row stands for the plot of $M_{++} (\left\{ W \right\},\tau)$ versus $\tau$ while the second one with $M_{-+} (\left\{ W \right\},\tau)$ versus $\tau$. The columns correspond to the repetition of the measurement for $3$ networks with $600$ nodes, $T_{s}=3$, $a=4$ and $p_{0}=0.001$ but distinct random initial positions of the nodes. The refractory periods are not identical and vary between $38$ and $40$. The ISI setpoint is fixed at $46$. The fraction of inhibitory nodes is $20 \%$. Red points correspond to the converged optimized network while the numerous blue crosses are associated with the random distribution of the weights among the network connections.}
\label{fig12}
\end{figure}

\begin{figure}
\resizebox{0.40\textwidth}{!}{
\includegraphics[]{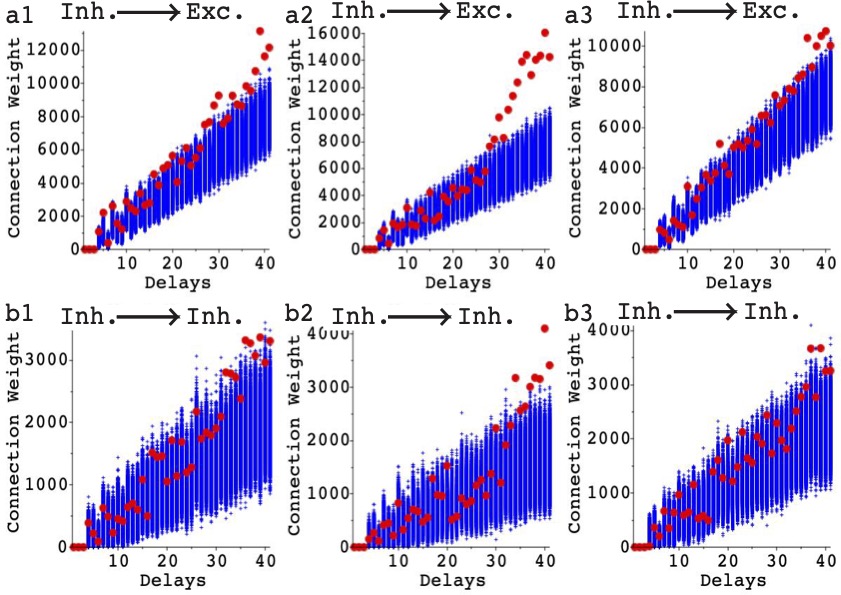}
}
\caption{Same regime of parameters as in fig.\ref{fig12}, but now the first row deals with $M_{+-}$  versus $\tau$ while the second one with $M_{--}$ versus $\tau$.}
\label{fig13}
\end{figure}

\begin{figure}
\resizebox{0.40\textwidth}{!}{
\includegraphics[]{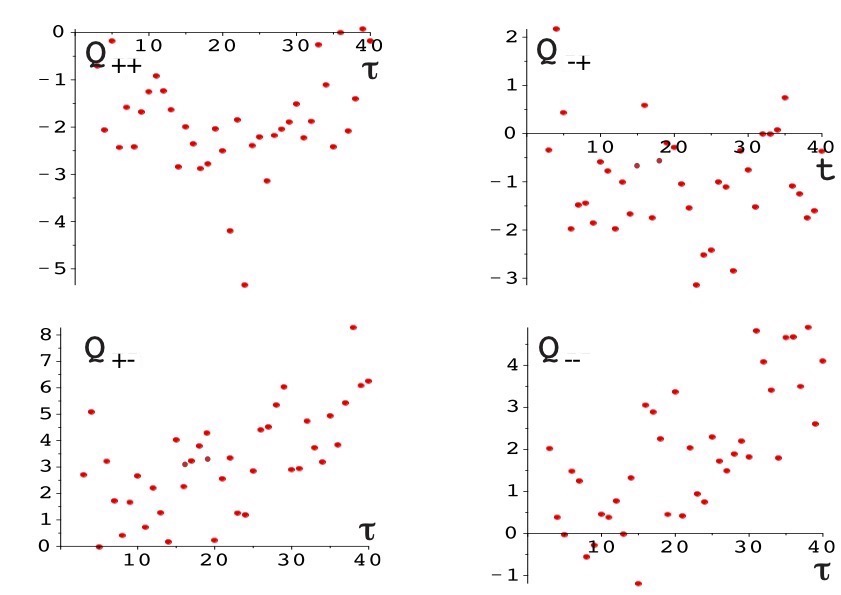}
}
\caption{Same regime of parameters as in the first column of fig.\ref{fig13} and fig.\ref{fig14}. The plots are concerned with $Q_{..}$ versus $\tau$ as defined in eq.\ref{definitionQ} which measures the deviation from the mean value in units of standard deviation.} 
\label{fig14}
\end{figure}

The question that interests us here is to know if there is a relationship between the propagation delay $\tau_{ij}$ between any 2 nodes $i$ and $j$ of the network and the weights $W_{ij}$ (possibly $W_{ji}$) of their connections. For that purpose, we introduce the following definitions:
\begin{equation}
M_{+-}(\left\{ W \right\},\tau)=\displaystyle{\sum_{\scriptsize{\begin{array}{l} \left\{i \in [1,N] \, \vert \, D_{i}=+1 \right\} \cr \left\{ j \in [1,N] \, \vert \, D_{j}=-1 \right\} \end{array}}}} \!\!\!\! \!\!\!\! \!\!\!\!\!\!\!\! W_{ij} \delta(\tau-\tau_{ij})
\end{equation}
 where $\delta(n)=1$ if $n=0$ and cancels out for any other integer value.  For the configuration $\left\{ W \right\}$, $M_{+-}(\left\{ W \right\},\tau)$ is the sum of the masses of all the connections from an inhibitory node to an excitatory one and separated by a propagation delay $\tau$. By analogy, we define in the same way $M_{++}$, $M_{-+}$ and $M_{--}$. We then proceed in the same way as for proving the anti-cluster feature of the optimized network or for proving the predominance of the inhibitory projections. We first compute $M_{..}(\left\{ W_{cvg} \right\},\tau)$ where $\left\{ W_{cvg} \right\}$ is the configuration network at the convergence of the optimization process and then we compare the result with $M_{..}(\left\{ W_{rand} \right\},\tau)$ where $\left\{ W_{rand} \right\}$ are derived from $\left\{ W_{cvg} \right\}$ by randomly redistributing its weights  among the nodes of the network.
 The results are displayed in the figures fig.\ref{fig12}, fig.\ref{fig13} and fig.\ref{fig14}. For fig.\ref{fig12}, fig.\ref{fig13}, the 3 columns correspond to the triple repetition of the numerical experiment by changing only the initial position of the nodes on the sphere. The red circles stand for the case of the optimized network $M_{..}(\left\{W_{cvg}\right\},\tau)$ while the numerous blue points are associated with $M_{..}(\left\{W_{rand}\right\},\tau)$ and the $10000$ random draw repetitions. Some figures give the impression that the optimized values are compatible with a random configuration of the connection masses. Others, on the contrary, seem to indicate that they clearly deviate from it. To clarify the situation, we introduce 
 \begin{equation}
 {\cal Q}_{-+}=
 { {M_{-+}(\left\{W_{cvg}\right\},\tau) - <\!\!M_{-+}(\left\{W_{rand}\right\},\tau)\!\!>}
 \over {\sqrt{<\!\!M_{-+}(\left\{W_{rand}\right\},\tau)^2\!\!>-<\!\!M_{-+}(\left\{W_{rand}\right\},\tau)\!\!>^2}} }
 \label{definitionQ}
 \end{equation}
 that stands for the deviation from the mean value measured in units of standard deviation (also ${\cal Q}_{++}$, ${\cal Q}_{+-}$ and ${\cal Q}_{--}$) 
 and plot it versus $\tau$ (fig.\ref{fig14}). The analysis of the figures leads to the following remarks:
 \begin{enumerate}
\item  For a given value of the delay, the values of $M_{++}$ or $M_{-+}$ associated with outgoing connections from excitatory nodes, do not deviate significantly (more than 3 standard deviations) from the mean value of the random distributions.
\item Nevertheless, if for a given delay, the values of $M_{.+}$ were only due to chance, then from one delay to another we should observe an alternation of values larger and smaller than the average. The fact that a large majority of the values are below the mean value is statistically significant and is corroborated by the global $P_{.+}$ measurements.
\item For outgoing connections from inhibitory nodes, we clearly observe that not only are $M_{.-}$ significantly above the random value, but also that this deviation increases with delay.
 The further the inhibitory connection projects, the higher its weight.
 \end{enumerate}
 
\section{Discussion}
We have just shown that a network of non-identical nodes, with excitable dynamics, pulse-coupled, with coupling delays depending on the Euclidean distance between nodes, was able to adapt the topology of its connections to obtain spike frequency synchronization.
The adapted network has the following remarkable properties:
\begin{enumerate}
\item The spatio-temporal dynamics is organized in a periodic succession of patterns. A pattern is constituted by the set of nodes that spiked at the same time. The set of patterns forms a partition of the network. There are very few connections between nodes of the same pattern and the vast majority of connections concern nodes belonging to distinct patterns. This results in an anti-cluster structure.
\item The network is very sparse. 
\item Inhibitory nodes play a fundamental role in frequency synchronization. Not only because frequency synchronization requires the presence of a minimum number of inhibitory nodes, but also because the total mass of outgoing connections from the inhibitory nodes is very significantly larger than if the connections were established randomly.
\item We observe the spontaneous occurrence of a spatial organization of inhibitory nodes: The further the inhibitory connection projects, the higher its weight.
\end{enumerate}

It is worth noting that these properties are somewhat generic in the sense that they do not depend on the details of the biological mechanisms that might have been involved. They derive solely from the fact that frequency synchronization has been imposed.

What could prevent our conclusions from applying to neural networks? First, although neuronal phase synchronization is suspected to play an important role in neuronal processes, the exact mechanism of operation remains to be discovered. Is phase synchronization a consequence of learning processes or a necessary prerequisite? Even more questionable is the willingness of neural networks to synchronize in frequency. 

Second, even if we admit the need for the neural network to be synchronized, there exist many biological mechanisms that could relax the geometric frustation character of the phase synchronization problem. For example, one can increase the duration of synaptic interaction, introduce mechanisms to adapt the spike frequency of an isolated neuron, or simply modify the propagation times of potentials by taking into account the myelic sheaths.

Finally and assuming that our results have any biological reality, the interpretation \cite{Izhikevich2006} of synchronization patterns in terms of information carriers is somewhat at odds with the anti-cluster structure that we observe. A good way to decide would be to measure the strength of synaptic coupling between neurons of the same pattern observed in experiments \cite{Ikegaya2004} and numerical simulations  \cite{Izhikevich2004} .

\end{document}